\documentclass{article}

\PassOptionsToPackage{numbers,compress}{natbib}

\usepackage[final]{neurips_2025}
\DeclareUnicodeCharacter{2192}{\ensuremath{\rightarrow}} 
\DeclareUnicodeCharacter{2264}{\ensuremath{\le}}         
\DeclareUnicodeCharacter{2265}{\ensuremath{\ge}}         

\usepackage[utf8]{inputenc}
\usepackage{caption}
\makeatletter
\providecommand{\@trackname}{}
\makeatother
\usepackage{caption}
\usepackage{amsmath}
\usepackage[T1]{fontenc}
\usepackage{hyperref}
\usepackage{tikz}
\usetikzlibrary{arrows.meta, positioning}
\usepackage{url}
\usepackage{booktabs}
\usepackage{amsmath,amssymb,amsfonts}
\usepackage{nicefrac}
\usepackage{microtype}
\usepackage{xcolor}

\usepackage{algorithm}
\usepackage{algpseudocode}

\usepackage{xpatch}
\makeatletter
\xapptocmd{\NAT@bibsetnum}{%
  \setlength{\leftmargin}{0pt}%
  \setlength{\itemindent}{\labelwidth}%
  \addtolength{\itemindent}{\labelsep}%
}{}{}
\makeatother

\newcommand{\Gtrain}{\mathcal{G}_{\mathrm{train}}}
\newcommand{\Gval}{\mathcal{G}_{\mathrm{val}}}
\newcommand{\TopM}{\mathrm{TopM}}
\newcommand{\simfn}{\mathrm{sim}}
\newcommand{\argmax}{\operatorname*{arg\,max}}
\newcommand{\argmin}{\operatorname*{arg\,min}}

\title{RAG-Enhanced Kernel-Based Heuristic Synthesis (RKHS):
A Structured Methodology Using Large Language Models for Hardware Design
}

\author{
Shiva Ahir \qquad Alex Doboli \\
Department of Electrical \& Computer Engineering \\
Stony Brook University, New York \\
\texttt{\{shiva.ahir, alex.doboli\}@stonybrook.edu}
}

\begin{document}
\maketitle

\begin{abstract}
Heuristic design upholds modern electronic design automation (EDA) tools, yet crafting effective placement, routing, and scheduling strategies entails substantial expertise. We study how large language models (LLMs) can \emph{systematically} synthesize reusable optimization heuristics beyond one-shot code generation \citep{ref2,ref3,ref4}. We propose \textbf{RAG-Enhanced Kernel-Based Heuristic Synthesis (RKHS)}, which integrates retrieval-augmented generation (RAG) \citep{ref1}, compact \emph{kernel} heuristic templates \citep{ref9,ref10}, and an LLM-driven refinement loop inspired by iterative self-feedback \citep{ref6}. Applied to latency-minimizing list scheduling in high-level synthesis (HLS) \citep{ref11,ref14}, a prototype reduces average schedule length by up to 11\% over a baseline scheduler with only 1.3$\times$ runtime overhead, and the structured retrieval-synthesis loop generalizes to other EDA optimization problems.
\end{abstract}

\section*{Introduction}
\addcontentsline{toc}{section}{Introduction}

Modern EDA relies on handcrafted heuristics to address NP-hard problems in placement, routing, scheduling, and high-level synthesis \citep{ref11,ref14}. These heuristics encode expert knowledge about structural patterns and performance–power–area (PPA) trade-offs, but their design is labor-intensive, domain-specific, and difficult to generalize across benchmarks or technology nodes. As design complexity increases, manual heuristic design and refinement become unsustainable.

Graph motif mining has been studied in network science to detect recurring subgraph structures and classical scheduling heuristics exploit structural signals such as critical paths and fanout \citep{ref11,ref14}. However, prior work does not systematically extract and cluster motifs from training DAGs for reusable heuristic synthesis. RKHS bridges this gap by deterministically mining structural motifs and integrating retrieved motif kernels into LLM-generated priority functions.


Most existing approaches treat LLMs as one-shot code generators rather than structured synthesizers of reusable optimization strategies for graph scheduling \citep{ref11,ref14}. They lack a principled mechanism to extract structural motifs from prior designs and recombine them into customized priority functions with consistent metric gains. We address this limitation by casting heuristic synthesis as a retrieval-augmented, kernel-based optimization loop, leveraging structural retrieval \citep{ref2} and reusable kernels grounded in kernel-method perspectives.


\setcounter{section}{1}

\subsection{Proposed Solution}

We focus on the following question: \textit{Can we automatically synthesize optimization heuristics that are executable, interpretable, and transferable across related instances of an EDA problem class?}
We treat list scheduling as a fixed \emph{meta-algorithm}, where the key design parameter is the priority function that ranks ready operations. Rather than redesigning the scheduler, we automatically customize this rule for each input graph. In contrast to reinforcement learning or black-box search methods that tune heuristic parameters globally, RKHS explicitly decomposes heuristic synthesis into structural motif retrieval and interpretable template integration. Unlike prior work that generates heuristics directly via one-shot LLM prompting~\cite{ref11,ref14}, we separate \emph{structural retrieval} from \emph{heuristic synthesis}: reusable motifs extracted from existing designs~\cite{ref9,ref10} guide the construction of a deterministic priority function grounded in kernel-based regularization~\cite{ref6}. This yields enhanced latency, interpretable scoring rules, and transferability across related designs through retrieval-driven adaptation.

\subsection{Problem Formulation}
We consider latency-minimizing list scheduling for high-level synthesis, a standard precedence-constrained scheduling problem with resource limits \citep{ref11,ref14}. The input is a directed acyclic graph (DAG) $G=(V,E)$ where each node $v\in V$ is an operation and each edge $(u,v)\in E$ encodes precedence $u \prec v$. Each operation has type $\tau(v)\in \mathcal{T}$ and duration $d(v)\in \mathbb{N}$ (often $d(v)=1$). Resource constraints are modeled by capacities $R_t\in\mathbb{N}$ for each type $t\in\mathcal{T}$, limiting the number of type-$t$ operations that can execute per cycle.

A schedule is a start-time assignment $s:V\rightarrow \mathbb{N}$ and is feasible if:
\begin{align}
\textbf{Precedence:}\quad
& s(v) \ge s(u) + d(u)
\qquad \forall (u,v)\in E
\label{eq:prec}
\\
\textbf{Resources:}\quad
&
\sum_{\substack{v\in V:\ \tau(v)=t}}
\mathbf{1}\{\, s(v)\le c < s(v)+d(v)\,\}
\ \le\ R_t
\qquad \forall t\in\mathcal{T},\ \forall c\in\mathbb{N}.
\label{eq:res}
\end{align}
The objective is to minimize makespan (latency):
\begin{equation}
L(s) \;=\; \max_{v\in V}\big(s(v)+d(v)\big),
\qquad
s^\star = \argmin_{s\ \text{feasible}} L(s).
\label{eq:obj}
\end{equation}

\paragraph{Baseline (Classical List Scheduling):}
List scheduling iteratively selects ready operations subject to resource constraints \citep{ref11,ref14}. Let $\mathrm{pred}(v)=\{u:(u,v)\in E\}$. At cycle $c$, the ready set is
\begin{equation}
\mathcal{R}(c)=
\left\{
v\in V\setminus \mathcal{S}:\ 
\forall u\in \mathrm{pred}(v),\ s(u)+d(u)\le c
\right\},
\label{eq:ready}
\end{equation}
where $\mathcal{S}$ is the set of already scheduled nodes. A list scheduler selects $\mathcal{A}(c)\subseteq \mathcal{R}(c)$ subject to resource limits and assigns $s(v)=c$ for $v\in\mathcal{A}(c)$. The key design freedom is the priority function $\pi(v;\,G,\mathrm{state})$, commonly based on levels, criticality, and fanout \citep{ref11,ref14}.


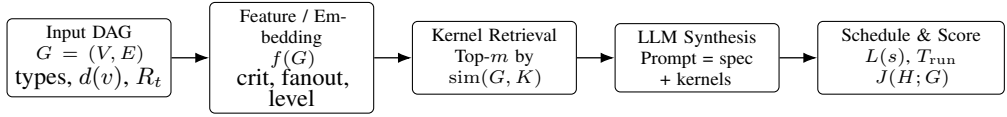
\begin{figure}[t]
\centering
\resizebox{0.95\linewidth}{!}{%
\begin{tikzpicture}[
  font=\scriptsize,
  box/.style={
    draw, rounded corners=2pt, align=center,
    inner sep=2.5pt, text width=20mm, minimum height=9mm
  },
  boxWide/.style={
    draw, rounded corners=2pt, align=center,
    inner sep=2.5pt, text width=23mm, minimum height=9mm
  },
  arr/.style={-Latex, line width=0.5pt}
]
\node[box] (in)
{Input DAG\\
$\displaystyle G=(V,E)$\\
\footnotesize types, $d(v)$, $R_t$};

\node[box, right=5mm of in] (feat)
{Feature / Embedding\\
$f(G)$\\
\footnotesize crit, fanout, level};

\node[box, right=5mm of feat] (retr)
{Kernel Retrieval\\
Top-$m$ by $\mathrm{sim}(G,K)$};

\node[box, right=5mm of retr] (llm)
{LLM Synthesis\\
Prompt = spec + kernels};

\node[boxWide, right=5mm of llm] (eval)
{Schedule \& Score\\
$L(s)$, $T_{\mathrm{run}}$\\
$J(H;G)$};

\draw[arr] (in) -- (feat);
\draw[arr] (feat) -- (retr);
\draw[arr] (retr) -- (llm);
\draw[arr] (llm) -- (eval);
\end{tikzpicture}%
}
\vspace{2mm}
\caption{\textbf{Overview of RKHS including feature extraction, retrieval, LLM synthesis, evaluation.}}
\label{fig:rkhs_overview}
\vspace{2mm}
\end{figure}

\subsection{RKHS: Solution Overview}

RKHS is a structured framework for synthesizing priority functions for resource-constrained list scheduling \citep{ref14,ref13}. RKHS treats heuristic design as a reusable pattern-learning problem: instead of generating a priority rule from scratch, it retrieves structurally similar scheduling patterns (kernels) distilled from training DAGs and uses them to guide synthesis as explained in Algorithm~1.

Each kernel captures two components: 
(i) a structural signature summarizing recurring motifs and graph statistics (e.g., critical-path summaries, fanout distributions, and reconvergence markers), and 
(ii) a heuristic template encoding scheduling preferences associated with those structures. Unlike classical kernel methods that operate in feature space, our kernels represent reusable structural scheduling primitives distilled from prior DAGs. The kernel library is constructed from training graphs via deterministic motif extraction and structural clustering (details in Appendix). At inference time, given a new graph $G$, we compute a deterministic structural embedding and retrieve the most similar kernels. Their structural signatures and templates are provided to the LLM via retrieval-augmented prompting, which synthesizes a compact, deterministic priority function $\pi(v)$. The synthesized rule is then evaluated using list scheduling under precedence and resource constraints, and refined using latency-based feedback \citep{ref6}.

The key novelty lies in (1) identifying reusable structural motifs from prior designs and (2) combining multiple retrieved kernel signals during synthesis, enabling priority functions that integrate complementary scheduling cues such as criticality, reconvergence, and resource pressure.



\subsection{Similarity and Retrieval}
Let $f(G)$ denote an embedding of a graph and $f(K)$ the embedding of a kernel signature. We use cosine similarity:
\begin{equation}
\simfn(G,K)=\frac{\langle f(G), f(K)\rangle}{\|f(G)\|_2\,\|f(K)\|_2}.
\label{eq:cosine}
\end{equation}
RKHS retrieves top-$m$ kernels:
\begin{equation}
\mathcal{K}_G = \TopM_{K\in\mathcal{K}}\ \simfn(G,K).
\label{eq:topm}
\end{equation}

\subsection{Heuristic Synthesis Objective}
RKHS seeks a parameterized heuristic policy $H_\theta$ that maps scheduler state and graph context to a priority score:
\begin{equation}
H_\theta:\ (\mathrm{state},G,v)\ \mapsto\ \pi(v),
\qquad v\in \mathcal{R}(c).
\label{eq:policy}
\end{equation}
The scheduler selects feasible actions greedily by descending priority, following standard list scheduling practice \citep{ref11,ref14}:
\begin{equation}
\mathcal{A}(c)=\textsc{SelectFeasible}\big(\mathcal{R}(c),\{\pi(v)\}_{v\in \mathcal{R}(c)}, \{R_t\}\big).
\label{eq:select}
\end{equation}

\paragraph{Scoring Function:}
Each candidate heuristic is evaluated by a score combining latency and runtime with infeasibility penalties:
\begin{equation}
J(H;G)
=
- L\big(s_H(G)\big)
- \lambda \, T_{\text{run}}(H,G)
- \mu \,\mathbf[\text{infeasible}],
\label{eq:exp_score}
\end{equation}
where $s_H(G)$ is the schedule produced by heuristic $H$ on graph $G$.

For a validation set $\Gval$, we report the mean validation score
\begin{equation}
\bar{J}(H) = \frac{1}{|\Gval|}
\sum_{G \in \Gval} J(H;G).
\label{eq:mean_score}
\end{equation}

\begin{algorithm}[t]
\captionsetup{labelformat=empty}
\caption{Algorithm 1: RAG-Enhanced Kernel-Based Heuristic Synthesis. Implementation: GPT-4, $|\mathcal{K}|=50$, $|\mathcal{G}_{\text{train}}|=200$, $|\mathcal{G}_{\text{val}}|=50$, $\lambda=0.01$, $\mu=5000$, $N=3$, $m=5$.}

\label{alg:rkhs}
\begin{algorithmic}[1]
\Require Training graphs $\Gtrain$, kernel library $\mathcal{K}$, LLM, iterations $N$, top-$m$ retrieval
\Ensure Best heuristic $H^\star$
\State Initialize history $\mathcal{H}\gets \emptyset$
\For{$i\gets 1$ \textbf{to} $N$}
    \State Sample batch $\mathcal{B}\subseteq \Gtrain$
    \ForAll{$G\in \mathcal{B}$}
        \State Retrieve kernels $\mathcal{K}_G \gets \TopM_{K\in\mathcal{K}}\, \simfn(G,K)$
        \Comment{Eq.~\eqref{eq:topm}}
    \EndFor
    \State Aggregate kernels $\mathcal{K}_{\mathcal{B}} \gets \bigcup_{G\in\mathcal{B}} \mathcal{K}_G$
    \State Construct prompt $P_i$ using spec, retrieved kernels $\mathcal{K}_{\mathcal{B}}$, constraints, and examples
    \State Query LLM to produce heuristic code $H_i$
    \Comment{LLM code generation: \citep{ref2,ref3,ref4}}
    \State Evaluate $H_i$ on $\Gval$ to obtain $\bar{J}_i$
    \Comment{Eq.~\eqref{eq:mean_score}}
    \State Append $(H_i,\bar{J}_i,\text{failure cases})$ to $\mathcal{H}$
    \State Generate feedback $F_i$ from failures/regressions; include $F_i$ in next prompt
    \Comment{Iterative refinement: \citep{ref6}; reasoning/acting: \citep{ref5}}
\EndFor
\State \Return $H^\star \gets \argmax_{(H,\bar{J})\in\mathcal{H}} \bar{J}$
\end{algorithmic}
\end{algorithm}

\subsection{RKHS Experiment Setup and Results}

\paragraph{Evaluation Protocol and Reporting:}
We report per-graph metrics as mean and sample standard deviation across validation graphs, and 95\% confidence interval half-widths assuming $n=|\Gval|$.

\paragraph{Baselines:}
We compare against a classical list scheduler using a level-based priority rule \citep{ref11,ref14}:
\[
\pi_{\text{base}}(v) = \text{level}(v)
\]
with deterministic ID tie-breaking.


\vspace{-6pt}

\begin{table}[h]
\centering
\small
\caption{RKHS synthesis loop: average metrics per iteration (higher $\bar{J}$ is better).}
\vspace{6pt}
\begin{tabular}{lccc}
\toprule
Iter & Latency $\downarrow$ & Runtime (ms) $\downarrow$ & $\bar{J}$ $\uparrow$ \\
\midrule
0 (Topo+ID)      & 65.05 & 82.53 & 63.79 \\
1 (Fanout-aware) & 49.45 & 89.78 & 74.59 \\
2 (Zero-in PQ)   & 75.85 & 87.93 & 71.89 \\
\bottomrule
\end{tabular}
\end{table}

\paragraph{Iteration-Level Results:}
Iteration 1 (fanout-aware intra-level prioritization) achieves the best composite score: the modest runtime increase is outweighed by a large latency reduction, consistent with known structural proxies for contention-heavy scheduling \citep{ref11,ref14}. Topo+ID = topological sort with ID ties, Fanout-aware = prioritize high-fanout nodes, Zero-in PQ = zero-slack priority queue.

\vspace{-17pt}


\begin{table}[h]
\centering
\small
\caption{Ablation study isolating the effects of retrieval and motif abstraction across representative validation graphs. Lower latency is better.}

\vspace{12pt}

\begin{tabular}{lcccc}
\toprule
Graph 
& Full RKHS 
& No Retrieval 
& No Motif 
& Random Kernel \\
\midrule
G1 & 52.0 & 60.3 & 63.5 & 68.1 \\
G2 & 98.8 & 109.4 & 112.0 & 120.5 \\
G3 & 95.2 & 104.7 & 108.2 & 115.9 \\
G4 & 71.8 & 82.5 & 86.4 & 91.3 \\
\midrule
Avg 
& 79.45 ($\pm$ 21.87)
& 89.23 ($\pm$ 22.57)
& 92.53 ($\pm$ 22.40)
& 98.95 ($\pm$ 24.23) \\
\bottomrule
\end{tabular}
\end{table}



We evaluated the RKHS prototype with targeted ablations to isolate retrieval and motif contributions. As shown in Table~2, the retained best heuristic achieves an average latency of 79.45 cycles across representative validation graphs. Removing similarity-based retrieval or motif abstraction increases both latency and variance, while random kernel selection further degrades performance. These results confirm that structurally grounded retrieval and deterministic motif extraction are essential for stable gains. Motif-level analysis shows that fanout motifs benefit parallel-heavy graphs, reconvergence motifs assist control-dense regions, and resource-aware motifs improve scheduling under tight operator constraints. The most stable latency is achieved by combining motif types, indicating that complementary structural signals enhance robustness. A comprehensive comparison with established state-of-the-art scheduling algorithms is part of ongoing work to assess scalability \& generalization.


\noindent\textbf{Interpretation:}
The retained policy can be expressed as a structure-aware priority:
\begin{equation}
\pi_{H^\star}(v)
=
\alpha \cdot \mathrm{crit}(v)
+
\beta \cdot \mathrm{fanout}(v)
-
\gamma \cdot \mathrm{level}(v),
\label{eq:best_policy}
\end{equation}
with deterministic tie-breaking. Here the \emph{remaining critical-path length} is defined as:
\begin{equation}
\mathrm{crit}(v)
=
\max_{p \in \mathcal{P}(v\rightarrow \mathrm{sink})}\ \sum_{u\in p} d(u),
\label{eq:crit_def}
\end{equation}
where $\mathcal{P}(v\rightarrow \mathrm{sink})$ is the set of all directed paths from $v$ to any sink node. Prioritizing high-fanout and critical nodes reduces downstream idle cycles, yielding up to 11\% latency improvement over the baseline while keeping runtime within \textbf{1.3$\times$}. RKHS encourages such compact rules by conditioning synthesis on retrieved structural kernels \citep{ref1} and refining via failure-driven feedback \citep{ref6}.

\vspace{-6pt}

\section*{Conclusion and Future Work}

We proposed RKHS, a structured method for developing executable scheduling heuristics with LLMs. By combining deterministic motif retrieval with LLM-guided synthesis, and an iterative synthesis-evaluation scheme, RKHS formalizes heuristic synthesis as a retrieval-guided search process. We obtain an optimized description of priority function heuristics, which minimizes latency variability over our validation graphs, with significant opportunities available through the integration of neural synthesis with deterministic scheduling constraints. Our directions for future work include extending the algorithm to larger benchmark suites, leveraging learned graph embeddings such as node2vec and GNNs to improve search efficiency, adding to the library of kernels used, and extending applicability to other EDA applications such as placement, floorplanning, and downstream synthesis-based optimization for latency-area-power optimization.


\appendix
\section{Appendix: Kernel Construction and Motif Extraction Details}

\subsection{Deterministic Retrieval Embedding}

For retrieval, we compute a deterministic embedding $f(G) \in \mathbb{R}^d$ by concatenating normalized structural statistics:
\[
f(G) =
\left[
\text{crit-path summary},
\text{fanout histogram},
\text{level histogram},
\text{op-type histogram},
\text{resource-pressure proxies}
\right].
\]

All features are normalized using z-score over $G_{\text{train}}$ to ensure reproducibility.

\subsection{Motif Identification}

Motifs are extracted deterministically from training DAGs by mining recurring labeled subgraphs such as:

\begin{itemize}
    \item $k$-hop neighborhoods
    \item High-centrality subgraphs
    \item Reconvergent fanout regions
    \item Deep linear chains
\end{itemize}

Each candidate subgraph is featurized using structural statistics and clustered by signature similarity.

\subsection{Reconvergence Marker Definition}

We quantify reconvergence at node $v$ as:
\[
R(v)=
\left|
\left\{
(u,w):
u,w \in \text{children}(v),
\exists x \text{ s.t. } u \rightarrow x \wedge w \rightarrow x
\right\}
\right|.
\]

High $R(v)$ indicates structural regions prone to downstream contention.

\subsection{Example Kernel Instantiations}
\label{sec:example_kernels}

\paragraph{Kernel A: Reconvergent Region}

\textbf{Structural Signature:}  
High reconvergence index $R(v)$, moderate fanout, and convergence of multiple child paths into shared downstream nodes. This pattern captures structural regions prone to contention and synchronization effects.

\textbf{Heuristic Template.}  
\[
\pi(v) =
\alpha_1 \cdot \text{crit}(v)
+
\alpha_2 \cdot R(v)
+
\alpha_3 \cdot \text{fanout}(v),
\]
where $\alpha_1, \alpha_2, \alpha_3$ are tunable weights.

\paragraph{Kernel B: Deep Critical Chain}

\textbf{Structural Signature:}  
High average critical-path depth with low slack values, indicating latency-sensitive chain-dominated regions.

\textbf{Heuristic Template.}  
\[
\pi(v) =
\beta_1 \cdot \text{crit}(v)
-
\beta_2 \cdot \text{slack}(v),
\]
where $\beta_1, \beta_2$ control the emphasis on urgency versus flexibility.

\subsection{Motif Combination}

When multiple kernels are retrieved, their structural signatures are jointly provided to the LLM. The synthesized priority function integrates complementary signals (e.g., criticality, reconvergence, resource pressure) into a unified scoring rule.

\section{Detailed Explanation of Algorithm 1}

\subsection{Overview}

Algorithm~1 describes the full RKHS synthesis loop. The goal is to generate a 
customized priority function $\pi(v)$ for a given DAG by combining:
(i) structural retrieval from a kernel library, and 
(ii) LLM-guided heuristic synthesis with evaluation feedback.

\subsection{Step-by-Step Explanation}

\textbf{Step 1: Initialization}

We initialize an empty history set $\mathcal{H}$ to store:
\begin{itemize}
    \item synthesized priority functions $H_i$,
    \item their latency scores $\bar{J}_i$,
    \item failure cases.
\end{itemize}

This history enables iterative refinement.

\medskip
\textbf{Step 2: Batch Sampling}

At each iteration, a batch of training graphs 
$\mathcal{B} \subseteq \mathcal{G}_{train}$ is sampled.

This prevents overfitting to a single graph and improves generalization.

\medskip
\textbf{Step 3: Kernel Retrieval}

For each graph $G \in \mathcal{B}$:

\begin{enumerate}
    \item Compute deterministic embedding $f(G)$.
    \item Compare against kernel embeddings $f(K)$.
    \item Retrieve top-$m$ kernels using cosine similarity:
    \[
    \text{sim}(G,K) = \frac{\langle f(G), f(K)\rangle}
    {\|f(G)\|_2 \|f(K)\|_2}
    \]
\end{enumerate}

Each retrieved kernel contains:
\begin{itemize}
    \item structural motif signature,
    \item associated heuristic template.
\end{itemize}

All retrieved kernels are aggregated into $K_{\mathcal{B}}$.

\medskip
\textbf{Step 4: Prompt Construction}

A structured prompt $P_i$ is constructed containing:
\begin{itemize}
    \item target graph statistics,
    \item retrieved kernel motifs,
    \item heuristic templates,
    \item scheduling constraints.
\end{itemize}

This forms the retrieval-augmented context for synthesis.

\medskip
\textbf{Step 5: LLM-Based Heuristic Synthesis}

The LLM generates executable Python code defining:

\[
\pi(v) = \text{priority}(v, \text{state}, G)
\]

The function must be:
\begin{itemize}
    \item deterministic,
    \item interpretable,
    \item compatible with list scheduling.
\end{itemize}

\medskip
\textbf{Step 6: Evaluation}

The synthesized priority function is evaluated using
resource-constrained list scheduling.

Latency (makespan) $\bar{J}_i$ is measured.

\medskip
\textbf{Step 7: Feedback and Refinement}

If failures or inefficiencies are detected:
\begin{itemize}
    \item failure cases are summarized,
    \item feedback signals $F_i$ are generated,
    \item feedback is injected into the next prompt.
\end{itemize}

This enables iterative improvement across $N$ iterations.

\medskip
\textbf{Step 8: Best Heuristic Selection}

After $N$ iterations:

\[
H^* = \arg\max_{H_i \in \mathcal{H}} \bar{J}_i
\]

The best-performing priority rule is returned.

\end{document}